\documentclass[11pt,preprint,showpacs]{revtex4}
\usepackage{graphicx}% Include figure files
\usepackage{dcolumn}% Align table columns on decimal point
\usepackage{bm}% bold math
\usepackage{amsmath}
\usepackage{amssymb}
\usepackage{amsfonts}

\begin{document}
%\pagestyle{plain}
%\draft

\title{The nn quasi-free nd breakup cross section: discrepancies 
to theory and implications on the $^1S_0$ nn force}

\author{H. Wita{\l}a}
\affiliation{M. Smoluchowski Institute of Physics, Jagiellonian
University, PL-30059 Krak\'ow, Poland}

\author{W. Gl\"ockle}
\affiliation{Institut f\"ur theoretische Physik II,
Ruhr-Universit\"at Bochum, D-44780 Bochum, Germany}

\date{\today}

\begin{abstract}
Large discrepancies between quasi-free neutron-neutron (nn) cross
section data from 
 neutron-deuteron (nd)  breakup and theoretical predictions based on
 standard nucleon-nucleon (NN) and three-nucleon (3N) forces are
 pointed out. The nn $^1S_0$  interaction 
is shown to be dominant in that configuration and has to be increased
  to bring theory and data into agreement. Using the next-to-leading 
order (NLO) $^1S_0$ interaction of chiral perturbation theory ($\chi$PT)
   we demonstrate  that the nn QFS cross section only slightly depends 
on changes of the nn scattering length but is very sensitive to variations
    of the effective range parameter. In order to account for 
the reported discrepancies one must decrease the nn effective range
 parameter by about $\approx 12 \% $ from its value implied by 
charge symmetry and charge independence of nuclear forces.
\end{abstract}


\maketitle \setcounter{page}{1}

\section{Introduction}
\label{one}

The knowledge of the NN interaction is fundamental for interpreting nuclear
phenomena. The proton-proton (pp) experiments provide a solid data 
basis \cite{nijm90,nijm93}, which restricts theoretical assumptions about the
strong part of the pp force. In case of the neutron-proton (np) 
 system this is only
true to a smaller extent. The partial wave analysis 
of the np data \cite{nijm93} relies on the assumption that the isospin $ t=1 $
piece can be taken over from the pp system and only the $ t=0 $ part 
is free in the adjustment to the data. 
 The lack of a free neutron
target forbids neutron-neutron (nn) experiments, therefore the information
on the nn interaction can be deduced only in an indirect way. To that
aim the best tool seems to be the study of
 the three-nucleon (3N) system composed from
two neutrons and the proton. It is simple enough to allow a rigorous
theoretical treatment, e.g. in the framework of Faddeev
equations \cite{physrep96}. The
neutron-deuteron  elastic scattering together with the neutron
induced deuteron breakup, supplemented with the triton properties,
offer a data basis which can be used to test properties of the nn force.
Especially the nd breakup process  with its rich set of configurations for free
three outgoing nucleons seems to be a powerfull tool to test the nuclear
Hamiltonian. By comparing theoretical predicitons to the nd breakup
data in different configurations  not only can
the present day models of two-nucleon (2N) interactions be tested, but
also effects of three-nucleon forces (3NF's) can be studied.

The nn quasi-free scattering (QFS) refers to a situation where  
the outgoing proton
 is at rest in the laboratory system. In the nd breakup also np QFS
  is possible. Here one of the neutrons is at rest while
 the second neutron together with the proton form a quasi-freely
 scattered pair.

The reported  nn QFS cross sections taken at
$E_n^{lab}=26$~MeV \cite{siepe2} and at $E_n^{lab}=25$~MeV
\cite{ruan1} overestimate the nd theory by $\approx 18 \%$.
  Surprisingly, when instead of the nn pair the np pair is quasi-freely
scattered, the theory follows
 nicely the np QFS cross section data  taken in the $E_n^{lab}=26$~MeV
 nd breakup measurement \cite{siepe2}.
 That good description of the np QFS cross section contrasts with the
drastic discrepancy between the theory and the  nn QFS cross section
data taken in the same experiment \cite{siepe2}.

We do not expect surprises in case of 
the pp QFS data \cite{raup1,przyb1,zejma1}, since 
the information of the rich set of pp data has been incorporated 
into the pp forces. In fact a recent analysis \cite{wit_coul} 
including the Coulomb force to pp QFS data  lead to a nice agreement,
while in previous  analysis \cite{raup1,przyb1,zejma1}
the Coulomb force  was not yet
included. Additional theoretical efforts to include 
all effects of the Coulomb force beyond 
the ones in \cite{wit_coul} are underway.

In section \ref{two} we exemplify the stability of the QFS cross sections 
against changes of modern nuclear forces. We also  demonstrate that below
 $\approx 30$~MeV the $^1S_0$ and $^3S_1$-$^3D_1$ NN force components 
dominate the  QFS cross sections.
In section \ref{three} we analyse the np as well as the nn QFS data from 
\cite{siepe2}  in terms of rigorous solutions of the 3N
 Faddeev equation  and discuss necessary changes in the $^1S_0$ nn 
force component to remove the  discrepancies in the nn QFS
  cross section. Thereby a  detailed study  is performed using  the 
next-to-leading order (NLO)  chiral NN force, composed of contact 
interactions and the one-pion exchange potential. It reveals that the 
effective range parameter  is decisive to reconcile theory and data. 
The outcome is discussed in section \ref{four} and further experimental
insights on the nn force are proposed. Finally we summarize 
 in section \ref{five}.

\section{ Stability and sensitivity studies}
\label{two}

\begin{figure}
\includegraphics[scale=0.8]{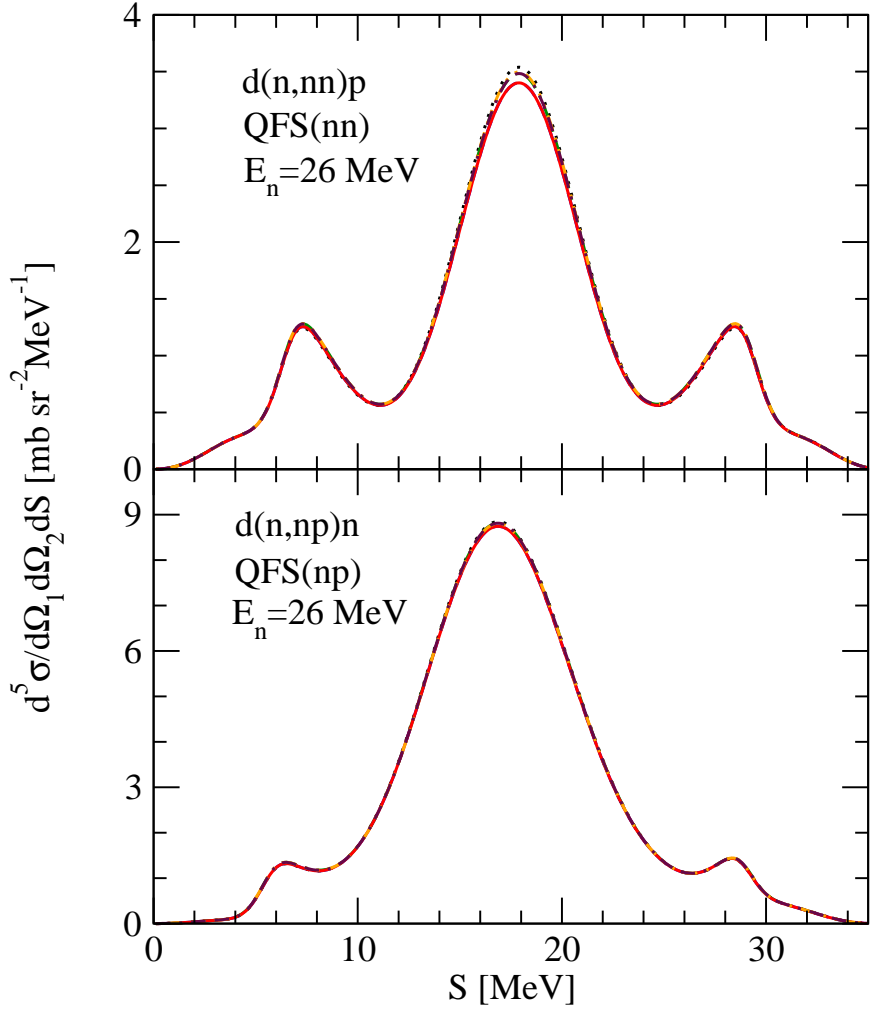}
\caption{(color online) The cross section $d^5\sigma/d\Omega_1d\Omega_2dS$
for the $E_n^{lab}=26$~MeV nd breakup reaction $d(n,nn)p$ (upper panel) and
$d(n,np)n$ (lower panel) as a function of the S-curve length for two
complete configurations of Ref.~\cite{siepe2}.
QFS nn refers to the angles of the two neutrons: $\theta_1=\theta_2=42^o$
 and QFS np refers to the angle $\theta_1=39^o$   of  the detected neutron
 and  $\theta_2=42^o$ for the proton. In both cases
 $\phi_{12}=180^o$.
The (practically overlapping)  lines
correspond to different underlying dynamics: CD Bonn~\cite{cdbonn}
 - dashed (blue), Nijm I
- dotted (black), Nijm II \cite{nijm}
 - dashed-dotted (green), CD Bonn+TM99 - solid
(red), Nijm I +TM99 \cite{TM,tm99}
 - dashed-double-dotted (orange), Nijm II + TM99 -
double-dashed-dotted (maroon). All partial waves with 2N total angular
momenta up to $j_{max}=5$ have been included.
}

\label{fig_qfs_nn_np_indep_NN_3NF}
\end{figure}

It is known   that nd  scattering theory provides  
QFS cross sections which are highly
independent from the realistic NN potential used in the calculations and
that they practically do not change when any of the present day 3NF's is
included~\cite{physrep96,kur2002,wit2010}. We exemplify it
in Fig.~\ref{fig_qfs_nn_np_indep_NN_3NF} for the nn and np QFS geometries 
of ref.~\cite{siepe2}. There results of 3N Faddeev calculations
\cite{physrep96} based on different high precision NN forces
(CD Bonn~\cite{cdbonn}, Nijm I and Nijm II~\cite{nijm}) alone or
combined with the TM99 3NF~\cite{TM,tm99} are shown.

\begin{figure}
\includegraphics[scale=0.8]{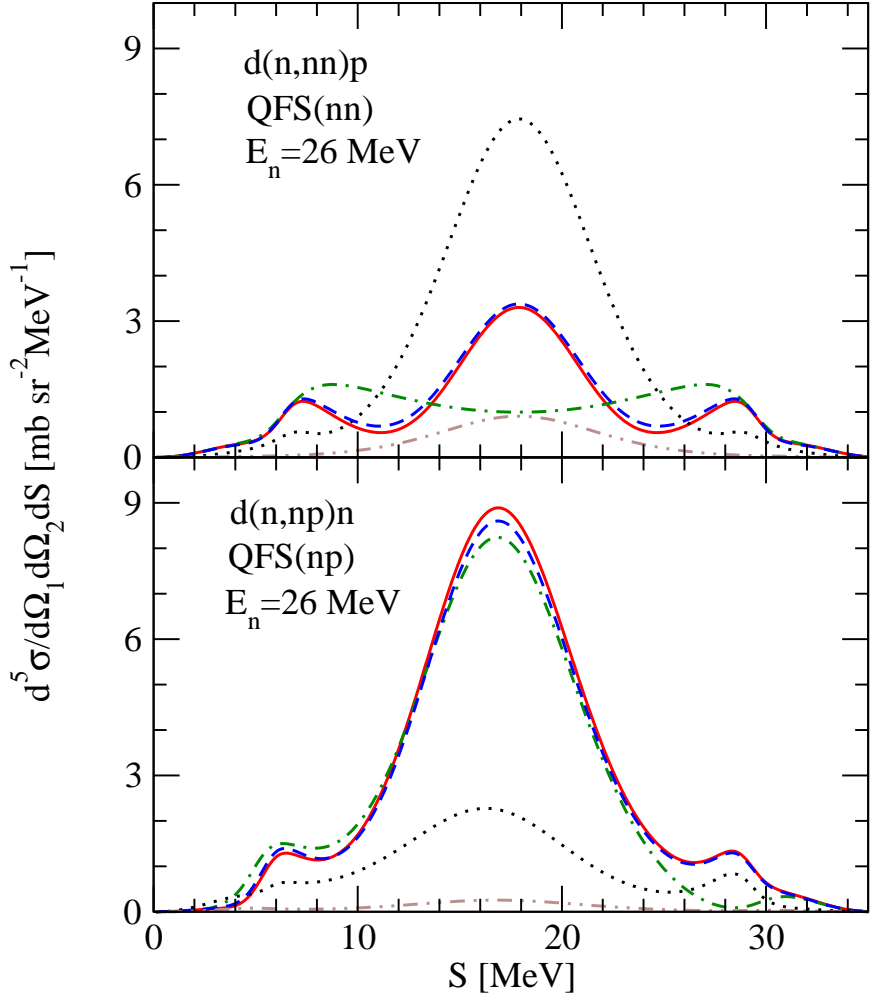}
\caption{(color online) The cross section $d^5\sigma/d\Omega_1d\Omega_2dS$
for the $E_n^{lab}=26$~MeV nd breakup reaction $d(n,nn)p$ (upper panel) and
$d(n,np)n$ (lower panel) as a function of the S-curve length for two
complete configurations of Ref.~\cite{siepe2} specified in
Fig.~\ref{fig_qfs_nn_np_indep_NN_3NF}.
 The different lines show contributions from
different NN force components. The solid (red) line is the full result based
on the CD Bonn potential \cite{cdbonn} and all partial waves with 2N total
angular momenta up to $j_{max}=5$ included. The dotted (black),
dashed-dotted (green), and dashed (blue) lines result when only
contributions from $^1S_0$, $^3S_1-^3D_1$, and $^1S_0+^3S_1-^3D_1$ are kept
calculating the cross sections. The dashed-double-dotted (brown) line
presents the contribution of all partial waves with the exception of $^1S_0$
and $^3S_1-^3D_1$.
}

\label{fig_sensiv_pw_qfsnnnp}
\end{figure}

The sensitivity study performed in \cite{wit2010}
revealed that at energies below $\approx 30$~MeV the $^1S_0$ and
$^3S_1$-$^3D_1$ NN force components provide the most dominant
contribution to the QFS cross sections with much less 
contributions of higher
partial waves. Specifically, in the  np QFS geometries the
$^3S_1$-$^3D_1$  is the dominant force component while for nn QFS it is the
$^1S_0$ force which contributes decisively. Again we exemplify it for
nn and np QFS geometries of ref.~\cite{siepe2} in
Fig.~\ref{fig_sensiv_pw_qfsnnnp}.

Such a dominance for the  QFS peak is understandable
since the QFS cross sections are practically insensitive to the action of 
the presently available 3NF. Then  at 
low energy the largest contribution should be provided by the
S-wave components of the NN potential. In case of  free np and nn 
scattering these are the
$^1S_0$(np)+$^3S_1-^3D_1$ and $^1S_0$(nn) contributions,
 respectively. In the simple minded spirit that under QFS condition 
one of the three nucleons (at rest in the lab system) is just a
  spectator such a dominance of a two-nucleon encounter is to be
  expected. 
In reality, however, the projectile nucleon also interacts  with
  that " spectator" particle and the three nucleons at low energies
  undergo 
higher order rescatterings \cite{physrep96,hwit89}.
 Thus the scattering to the final nn (np) QFS configuration also 
receives contributions from the np $ ^3S_1-^3D_1 $ (nn $ ^1S_0 $) interaction.
  Despite of all that the numerical results clearly reveal that for 
the np QFS configuration the $ ^3S_1-^3D_1 $ force is the most
dominant 
contribution and for the nn QFS it is the $  ^1S_0 $ force (for free
nn scattering there is no  $ ^3S_1-^3D_1 $ interaction possible). 
This implies that the nn QFS is a powerful
   tool to study the $  ^1S_0 $  nn force component.

\begin{figure}
\includegraphics[scale=0.9]{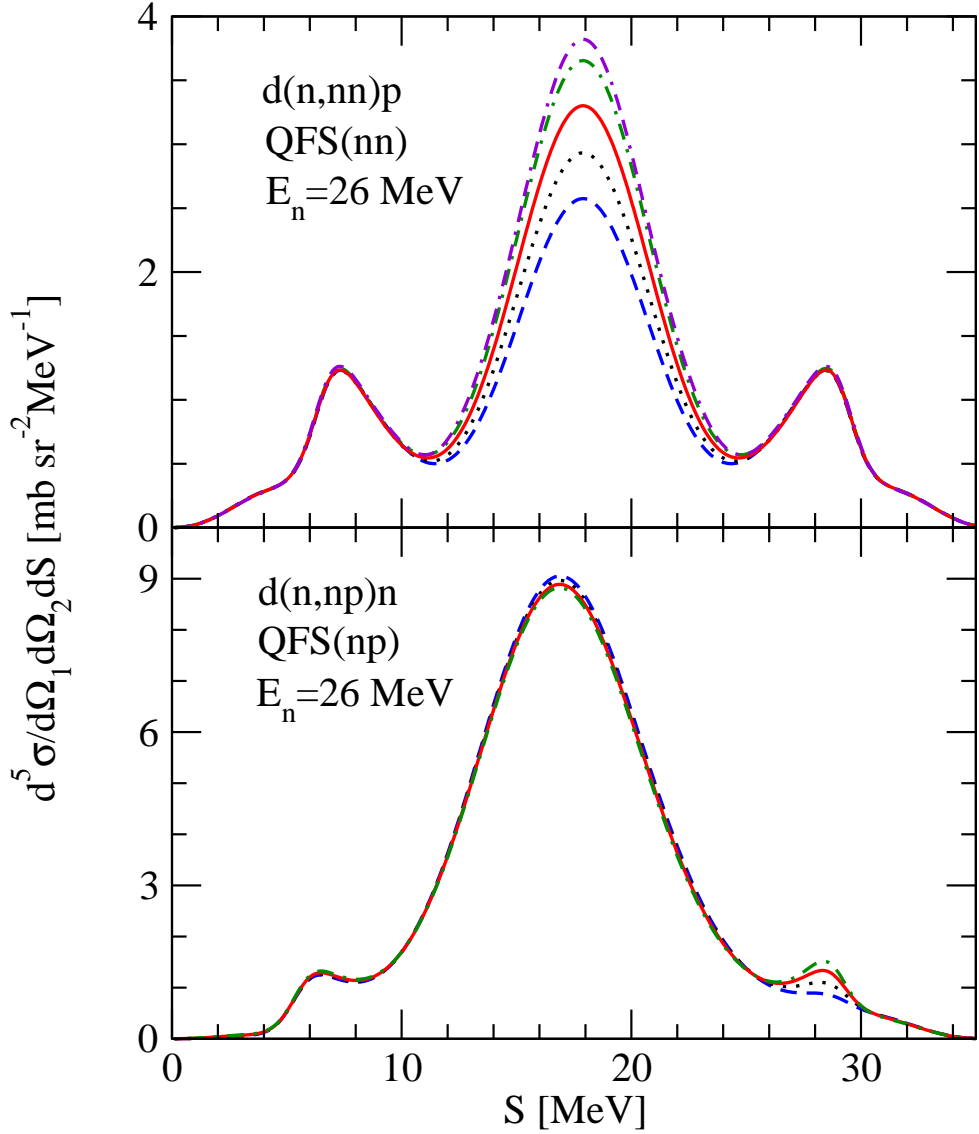}
\caption{(color online) The cross section $d^5\sigma/d\Omega_1d\Omega_2dS$
for the $E_n^{lab}=26$~MeV nd breakup reaction $d(n,nn)p$ (upper panel) and
$d(n,np)n$ (lower panel) as a function of the S-curve length for two
complete configurations of Ref.~\cite{siepe2} specified in 
Fig.~\ref{fig_qfs_nn_np_indep_NN_3NF}.
 The
lines show sensitivity of the QFS cross sections to the changes of the nn
$^1S_0$ force component. Those changes were induced by multiplying the
$^1S_0$ nn matrix element of the CD Bonn potential by a factor $\lambda$.
The solid (red) line is the full result based on the original CD Bonn
potential \cite{cdbonn} 
($a_{nn}=-18.8$~fm, $r_{eff}=2.79$~fm) 
 and all partial waves with 2N total angular momenta
up to $j_{max}=5$ included. The dashed (blue), dotted (black), and
dashed-dotted (green) lines correspond to 
$\lambda=0.9$ ($a_{nn}=-8.3$~fm, $r_{eff}=3.12$~fm),
 $0.95$  ($a_{nn}=-11.7$~fm, $r_{eff}=2.96$~fm),
and $1.05$ ($a_{nn}=-42.0$~fm, $r_{eff}=2.66$~fm),
respectively. The double-dashed-dotted (violet)
 line shows cross sections obtained with  
$\lambda=1.08$ ($a_{nn}=-134.7$~fm, $r_{eff}=2.61$~fm), which factor is
required to get agreement with nn QFS data of
ref. \cite{siepe2}. 
}
\label{fig_1s0nn_changes_qfs}
\end{figure}

That extreme sensitivity of the nn QFS cross section to the $^1S_0 $  
nn force component is demonstrated in Fig.~\ref{fig_1s0nn_changes_qfs} 
 for the QFS geometries of ref.~\cite{siepe2}.  
To that aim we multiplied the $  ^1S_0 $
  nn matrix element of the CD Bonn potential by a factor $ \lambda$. 
The result  is, that  the nn QFS cross section  undergoes
 significant variations while the np QFS cross section is practically
 unchanged.  The displayed $\lambda$-parameters include also the
 value $ \lambda=1.08 $ 
 which is necessary to get agreement with the nn QFS data of
 ref.\cite{siepe2}.

\begin{figure}
\includegraphics[scale=0.89]{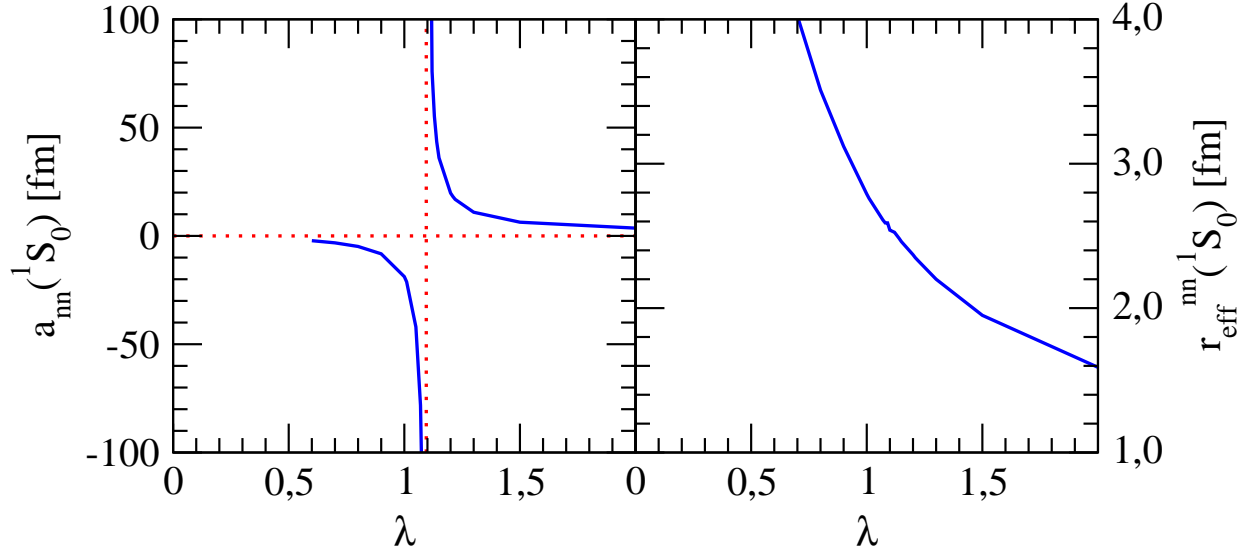}
\caption{(color online) The changes of the nn scattering length
  $a_{nn}$ and the effective range parameter $r_{eff}$ with factor $\lambda$ by
  which the $^1S_0$ nn matrix element of the CD Bonn
potential is multiplied: $V_{nn}(^1S_0)=\lambda * V_{CD~Bonn}(^1S_0)$.
}
\label{ann_reff_fac}
\end{figure}

\begin{figure}
\includegraphics[scale=0.9]{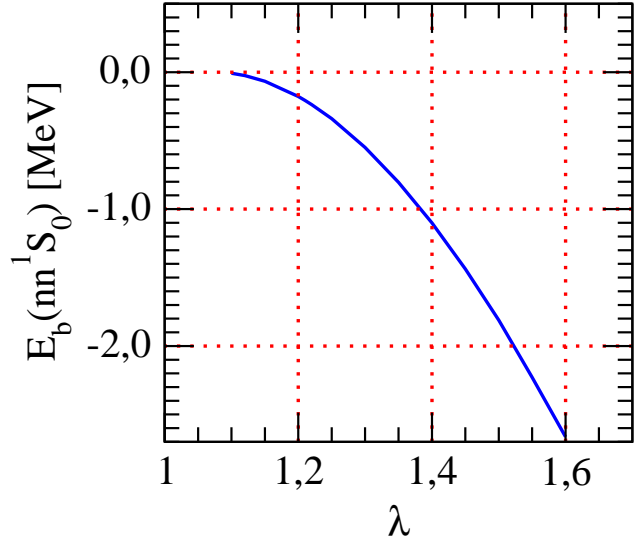}
\caption{(color online) The range of $\lambda$ values  by
  which the $^1S_0$ nn matrix element of the CD Bonn
potential is multiplied ($V_{nn}(^1S_0)=\lambda *
V_{CD~Bonn}(^1S_0)$), for  which  the two neutrons form a bound state with
 the binding energy $E_b$.
}
\label{ebind1s0_fac}
\end{figure}

While both,  $^1S_0$  and $^3S_1-^3D_1$, np forces are well
determined by np scattering data (with the restrictions mentioned
above) and by the deuteron properties, the
$^1S_0$ nn force is determined up to now only indirectly due to lack
of free nn data.
 The disagreement between data and theory in the nn QFS peak
 points to the  possibility of a flaw in the nn $^1S_0$ force. It was 
shown in \cite{wit2010} that in order to remove the
$\approx 18 \%$ discrepancy found in \cite{siepe2} for  the nn QFS 
cross section
 required an increased strength of the   $^1S_0$ nn
interaction which when given in terms of a factor $\lambda$ amounts to
  $\lambda \approx 1.08$.
 In Fig.~\ref{ann_reff_fac}  we show the effect
of the $ \lambda$-modification for the nn scattering length $a_{nn}$
and 
for the  effective range
  parameter $r_{eff}$, and in Fig.~\ref{ebind1s0_fac} for the
 binding energy of two neutrons in  the $^1S_0$ state.
It is seen that taking $\lambda =1.08$   leads  to a nearly  bound
 state of two neutrons.

\section{ Implications on the $^1S_0$  nn effective range parameter}
\label{three}

Since the multiplication of the $^1S_0$ potential matrix element by a
factor $\lambda$ induces changes in the effective range as well as in
the scattering length
the question arises, which from both effects is more important for the nn
QFS cross section variations ? To answer that question  we
 performed 3N Faddeev calculations based on the
 next-to-leading (NLO) order  $\chi$PT potential \cite{epel2000,epel}
 including all np and nn forces up to the total angular momentum
 $j_{max}=3$ in the two-nucleon subsystem. The $^1S_0$ component of
 that interaction is composed of the one-pion exchange potential 
and contact interactions  parametrized by
 two parameters $\tilde C_{^1S_0}$ and $C_{^1S_0}$
\begin{equation}
V(^1S_0) = \tilde C_{^1S_0} + C_{^1S_0} (p^2 + p'^2) ~.
\end{equation}
Standard values are  $\tilde C_{^1S_0} =-0.1557374 *
10000$~GeV$^{-2}$
 and $C_{^1S_0} = 1.5075220 * 10000$~GeV$^{-4}$ for
cut-off combinations $\{\Lambda, \tilde {\Lambda} \} = \{450$~MeV
, $500$~MeV\} \cite{epel}.

Multiplying $\tilde C_{^1S_0}$ by a factor $C_2(^1S_0)$ and
$C_{^1S_0}$ by a factor $C_1(^1S_0)$  one can induce  changes
of the nn $^1S_0$ interaction. Requiring either the value of the scattering
length $a_{nn}$ or the value of the effective range parameter
$r_{eff}$  to be constant correlates  the  $C_1(^1S_0)$ and
$C_2(^1S_0)$ factors.

\begin{figure}
\includegraphics[scale=0.9]{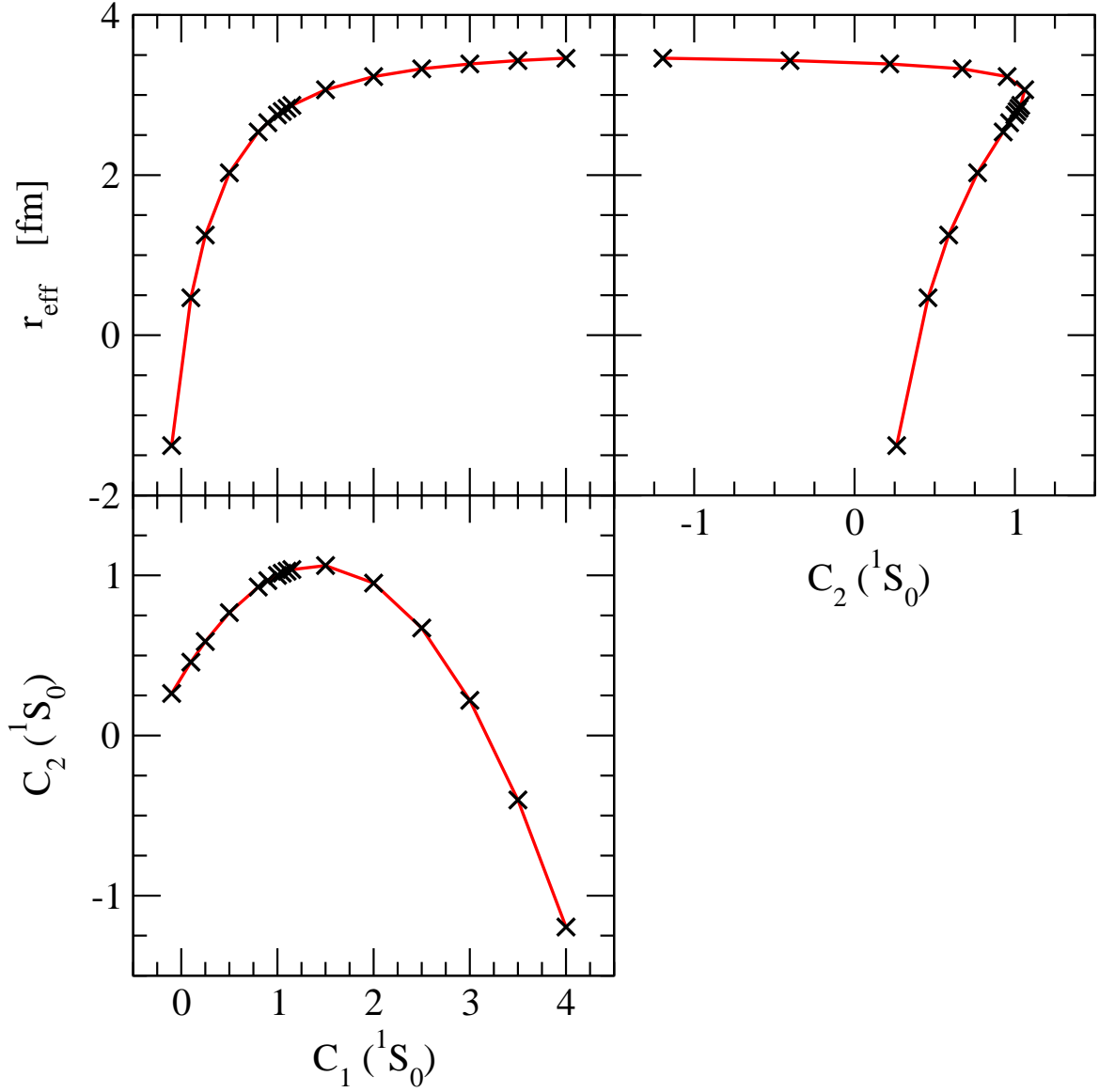}
\caption{(color online) Changes of the effective range parameter
  $r_{eff}$ in the
  $^1S_0$ partial wave caused by a correlated change of the factors
  $C_1(^1S_0)$ and  $C_2(^1S_0)$ as shown in bottom part of this figure.
This correlation between the factors  $C_1(^1S_0)$ and  $C_2(^1S_0)$
corresponds to a constant value of the scattering length
$a_{nn}=-17.6$~fm.
}
\label{reff_from_c1_c2_new}
\end{figure}

\begin{figure}
\includegraphics[scale=0.9]{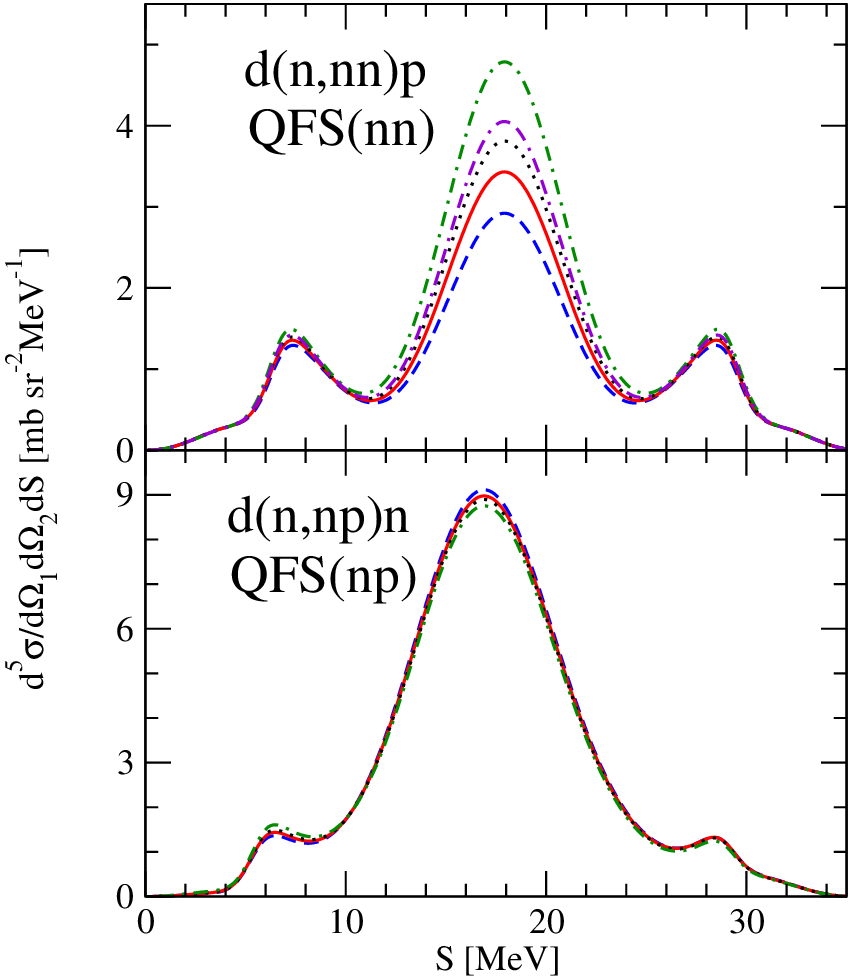}
\caption{(color online) Changes of QFS cross sections  
 for configurations 
specified in Fig.~\ref{fig_qfs_nn_np_indep_NN_3NF} 
caused by 
 correlated change of factors
  $C_1(^1S_0)$ and  $C_2(^1S_0)$ shown  in
 Fig.~\ref{reff_from_c1_c2_new}.  All lines show results of Faddeev
 calculations based on LO $\chi$PT potential and
 all partial waves with 2N total
angular momenta up to $j_{max}=3$ included. They differ in the nn $^1S_0$
force which was obtained keeping constant scattering length
$a_{nn}=-17.6$~fm and changing constants $C_1(^1S_0)$ and
$C_2(^1S_0)$ to get different effective ranges which are: solid (red
line) - $C_1(^1S_0)=1.0$, $C_2(^1S_0)=1.0$, $r_{eff}=2.75$~fm, 
dashed (blue line) - $C_1(^1S_0)=1.5$, $C_2(^1S_0)=1.0615$,
$r_{eff}=3.07$~fm, 
dotted (black line) - $C_1(^1S_0)=0.8$, $C_2(^1S_0)=0.9275$,
$r_{eff}=2.54$~fm, 
dashed-dotted (green line) - $C_1(^1S_0)=0.5$, $C_2(^1S_0)=0.7675$,
$r_{eff}=2.03$~fm. 
The double-dashed-dotted (violet)
 line shows cross sections obtained with $C_1(^1S_0)=0.7064$, 
$C_2(^1S_0)=0.8842$, $r_{eff}=2.41$~fm, which are 
required to get agreement with nn QFS data of
ref.\cite{siepe2}.
}
\label{e26p0_qfs_NLO_reff}
\end{figure}

Changing $C_1({^1S_0})$  and $C_2(^1S_0)$ in such a way that the
 scattering length is kept constant and equal $a_{nn}=-17.6$~fm leads to changes
 of the effective range $r_{eff}$ shown in
 Fig.~\ref{reff_from_c1_c2_new}.
 The resulting changes of the  nn and np QFS cross sections
  for  geometries of ref. \cite{siepe2}
 are shown in Fig.~\ref{e26p0_qfs_NLO_reff} for five  sets of $C_1(^1S_0)$ and
$C_2(^1S_0)$ factors with different nn $^1S_0$ effective range
 parameters ranging from $r_{eff}=2.03$~fm to $r_{eff}=3.07$~fm; one
 of them corresponding to the value required by the data.

\begin{figure}
\includegraphics[scale=0.9]{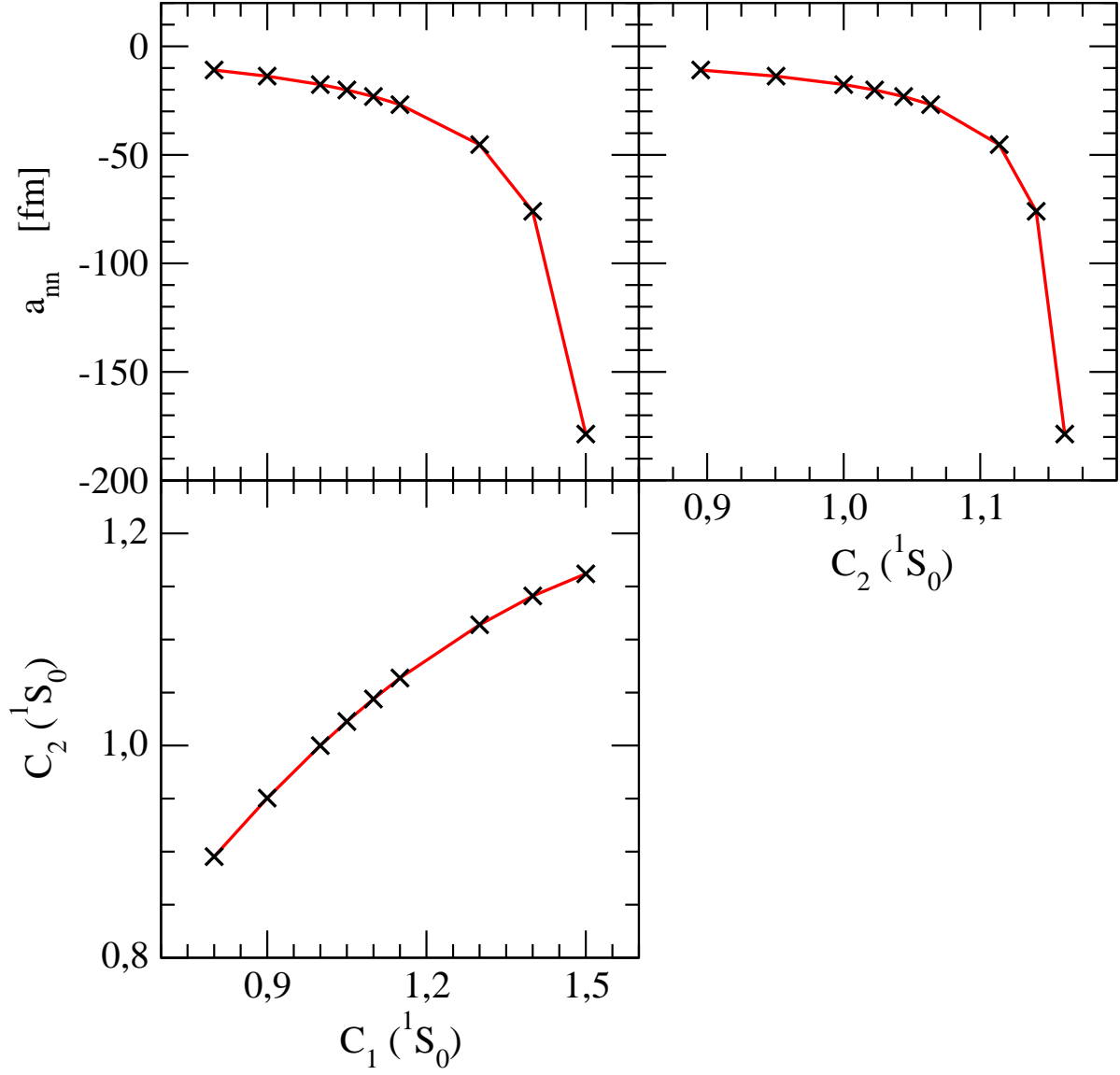}
\caption{(color online) Changes of the nn scattering length $a_{nn}$
 in the
  $^1S_0$ partial wave caused by a correlated change of the factors
  $C_1(^1S_0)$ and  $C_2(^1S_0)$ as shown in the bottom part of this figure.
This correlation between the factors  $C_1(^1S_0)$ and  $C_2(^1S_0)$
corresponds to a constant value of the effective range parameter
$r_{eff}=2.75$~fm.
}
\label{ann_from_c1_c2_new}
\end{figure}

\begin{figure}
\includegraphics[scale=0.9]{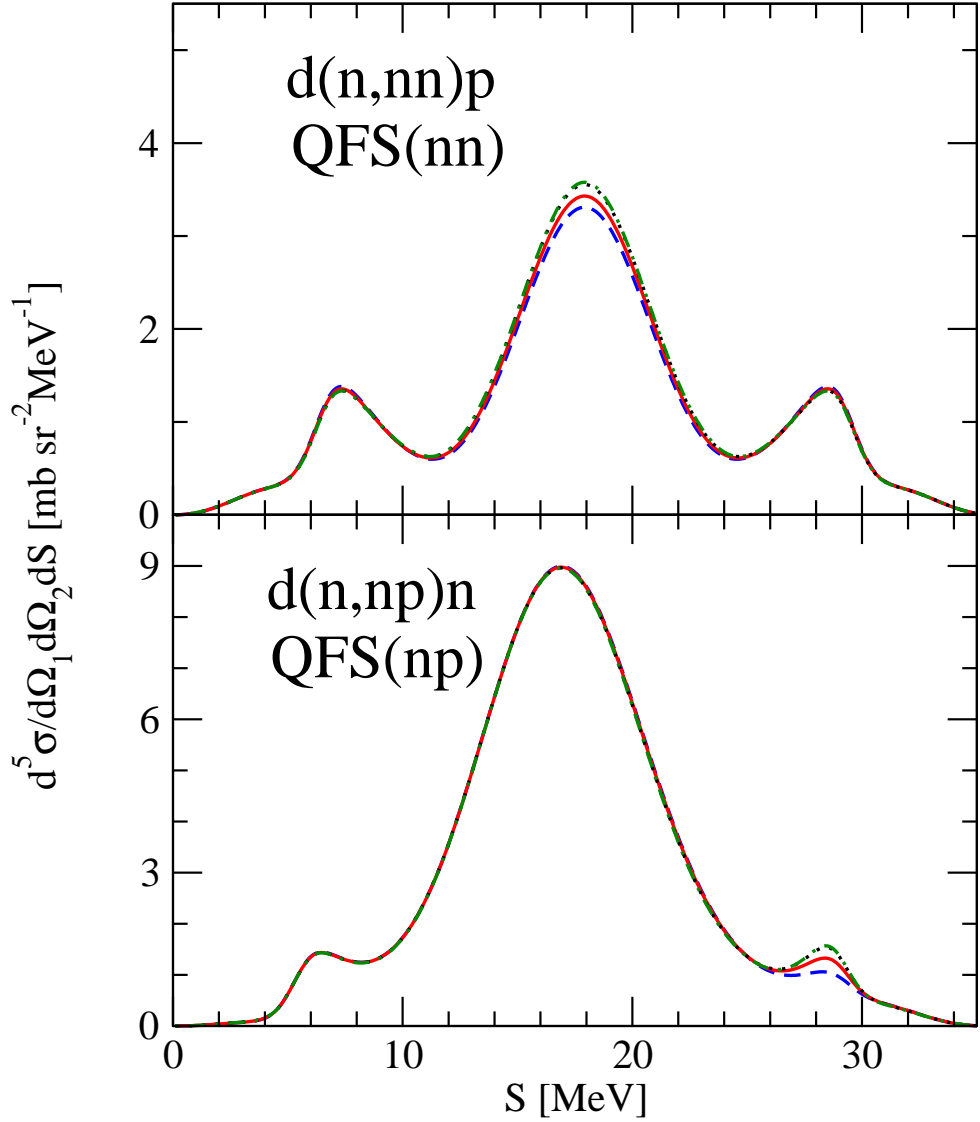}
\caption{(color online) Changes of QFS cross sections
 for configurations
specified in Fig.~\ref{fig_qfs_nn_np_indep_NN_3NF}
caused by
 a correlated change of the factors
  $C_1(^1S_0)$ and  $C_2(^1S_0)$ shown  in
 Fig.~\ref{ann_from_c1_c2_new}.
 All lines show results of Faddeev
 calculations based on the NLO $\chi$PT potential and
 all partial waves with 2N total
angular momenta up to $j_{max}=3$ included. They differ in the nn $^1S_0$
force which was obtained keeping the  effective range parameter
$r_{eff}=2.75$~fm constant and changing the constants $C_1(^1S_0)$ and
$C_2(^1S_0)$ to get different scattering lengths which are: solid (red
line) - $C_1(^1S_0)=1.0$, $C_2(^1S_0)=1.0$, $a_{nn}=-17.6$~fm,
dashed (blue line) - $C_1(^1S_0)=0.8$, $C_2(^1S_0)=0.8953$,
$a_{nn}=-10.9$~fm,
dotted (black line) - $C_1(^1S_0)=1.3$, $C_2(^1S_0)=1.1139$,
$a_{nn}=-45.3$~fm,
dashed-dotted (green line) - $C_1(^1S_0)=1.4$, $C_2(^1S_0)=1.1410$,
$a_{nn}=-76.0$~fm.
}
\label{e26p0_qfs_NLO_ann}
\end{figure}

Similarily, changing  $C_1(^1S_0)$  and $C_2(^1S_0)$ while keeping
 the effective range  constant to $r_{eff}=2.75$~fm,
  leads to changes of the
  nn $^1S_0$ scattering length $a_{nn}$ shown in
 Fig.~\ref{ann_from_c1_c2_new}. The resulting changes of the nn and
 np QFS cross sections are presented in Fig.~\ref{e26p0_qfs_NLO_ann}
 for four values of the nn $^1S_0$ scattering length ranging from
 $a_{nn}=-10.9$~fm to $a_{nn}=-75.9$~fm. It
 is clearly seen that the nn QFS cross sections depend only slightly on
 a change of the scattering length. The variations of the QFS cross
 section 
maximum stays below  $\approx \pm 4 \%$. On the other side much
 stronger variations of the nn QFS cross sections result from changes
 of the  effective range (see  Fig.~\ref{e26p0_qfs_NLO_reff}).

Thus we can conclude  that the $\lambda$-enhancement  mechanism 
for the $^1S_0$ nn force studied in \cite{wit2010} acts mainly 
 through the change
 of the effective range parameter.
 Thus in order to remove the discrepancies found in \cite{siepe2} and
 \cite{ruan1} for the nn QFS cross section a
 change of the nn $^1S_0$ effective range parameter is required. Its  value
  taken under  the assumption of charge symmetry and charge
  independence of nuclear forces
  is $r_{eff}=2.75$~fm and it has to be changed  to $r_{eff} \approx
  2.41$~fm. 
That  implies a large charge symmetry and charge independence breaking 
 effect of about  $\approx 12 \%$
  for that parameter.

We would like to add that the discussed changes of $r_{eff}$ did not 
affect the elastic nd cross section nor vector or tensor analysing
powers to a measurable extent. 
 Only more complicated spin observables  in elastic nd scattering are 
affected but the present day experimental errors are much larger than 
those changes.

\section{ Discussion and further experimental information}
\label{four}

Is such a large isospin breaking effect at all possible in view of the present
understanding of nuclear forces?

First of all it seems unprobable that only the effective range would
reveal large isospin breaking  
 and the scattering length will be left unaffected.
In $\chi$PT the
leading  isospin breaking  contribution is provided  by  isospin
breaking  contact interaction without
derivatives \cite{epel_privat}. It turns out that the effective range
parameter is
  quite insensitive to that  isospin breaking  
 contact force and typical  isospin breaking  effects
  for $r_{eff}$ are small and under $\approx 1 \%$
  \cite{epel_privat}.

The reported
  discrepancies for nn QFS require however a much larger effect for $r_{eff}$ of
  the order $\approx 12 \%$. Only when the contact terms in next
  orders would be unnaturally  large one could expect larger  isospin
  breaking  effects 
  for $r_{eff}$. Assuming naturalness it seems rather unprobable.

Since it seems unlikely that  isospin breaking 
 effects will show up, if at all,  in
 the effective range parameter alone
without affecting simultaneously
the  nn scattering length, the question of a  possible
existence of a bound state of two neutrons reappears.

Present day NN interactions allow only one bound state of two
nucleons, namely the deuteron, where the neutron and the proton are
interacting in a state with angular momenta $l=0$ or $2$, total spin
$s=1$, and total angular momentum $j=1$. When the neutron and proton
are interacting with  the $^1S_0$ force  no bound state exists
and only a virtual resonant state occurs as documented by
the negative scattering length $a_{np}=-21.73$~fm.
 Also the data for the proton-proton system  exclude a $^1S_0$ pp
 bound state; however in this case the nuclear force is overpowered by
 the strong pp Coulomb repulsion.
 Assuming  charge-independence and charge-symmetry of strong
interactions also the two neutrons should not bind in the $^1S_0$ state.

It also seems that modern nuclear forces do not allow for the 3n and 4n
 systems to be bound \cite{pieper}.
However, in view of the strong discrepancies between theory and data
 found in the nd breakup measurements for the  nn QFS geometry,
which cannot be explained by
 present day nuclear forces, it appears  reasonable to check experimentally
 the possibility  of two neutrons being bound.

There are reactions which provide conditions advantegous for
a hypothetical di-neutron bound state.
  Such  conditions can be found e.g. when two
neutrons are moving with equal momenta and with relative energy close to
zero. That occurs in the so called final-state-interaction (FSI)
geometry of the nd breakup. Incomplete nd breakup measurements 
have been performed in the past to
study properties of the $^1S_0$ nn force \cite{tornow96}.
 Even a  dedicated experiment
 was performed in order to look for a hypothetical  $^1S_0$
nn bound state \cite{vwitsch2006} in which the spectrum of the proton
going in  forward direction
has been measured with the aim of a precise determination of its 
high energy region.
 The negative
result of \cite{vwitsch2006}
showed that the nd reaction is not suitable for such a study.

It seems that much more appropriate would be reactions  in which from
the begining two neutrons occupy a configuration advantegous  for
their binding.

\begin{figure}
\includegraphics[scale=0.8]{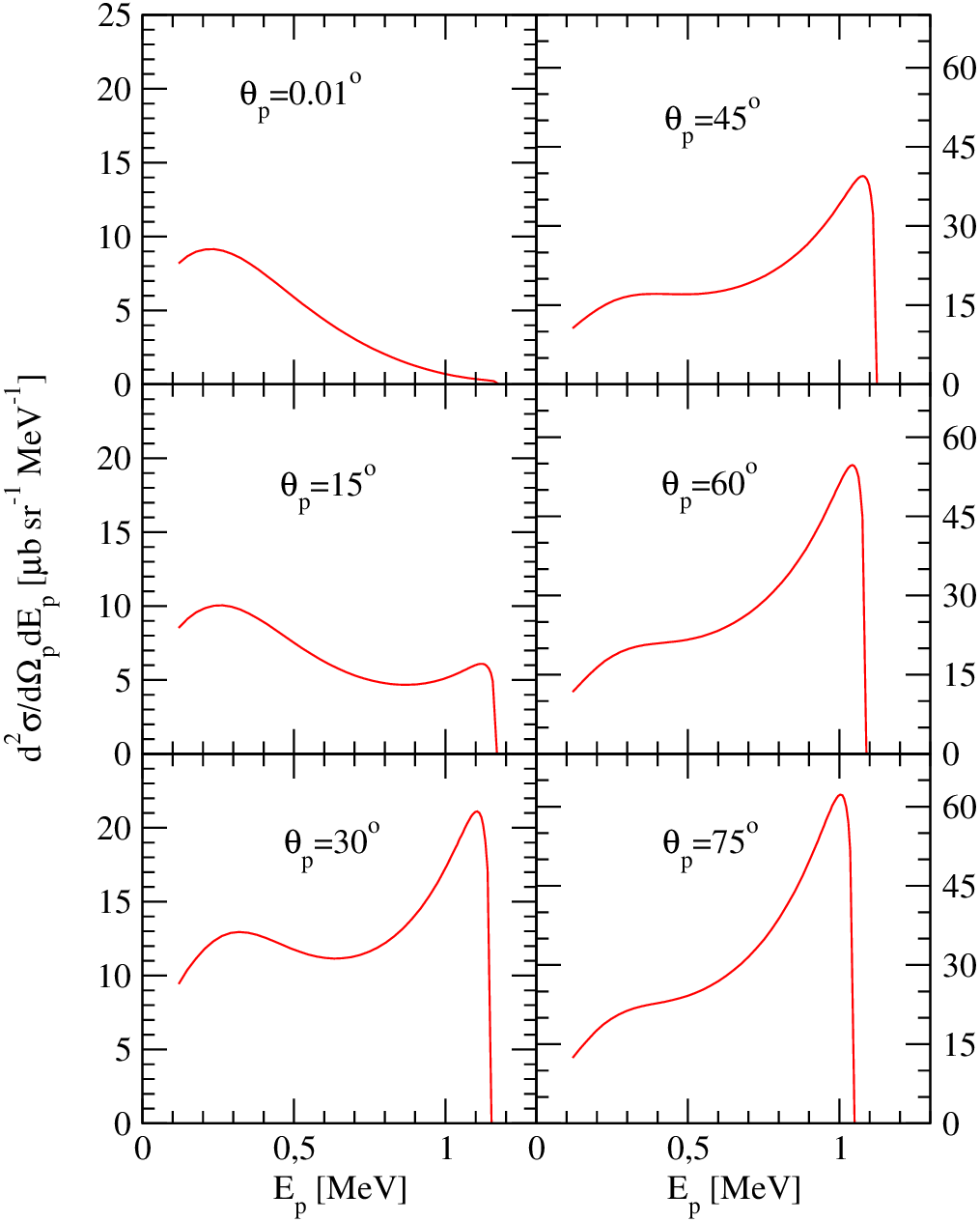}
\caption{(color online) The spectra of the outgoing proton from the
  reaction $^3H(\gamma,p)nn$ with $E_{\gamma}=10$~MeV at different
  lab.  angles of the proton. They have been calculated using
  the AV18~\cite{av18}
  NN interaction and the current composed of single nucleon and meson
  exchange currents~\cite{physrep2005}.
}
\label{fig1}
\end{figure}

\begin{figure}
\includegraphics[scale=0.8]{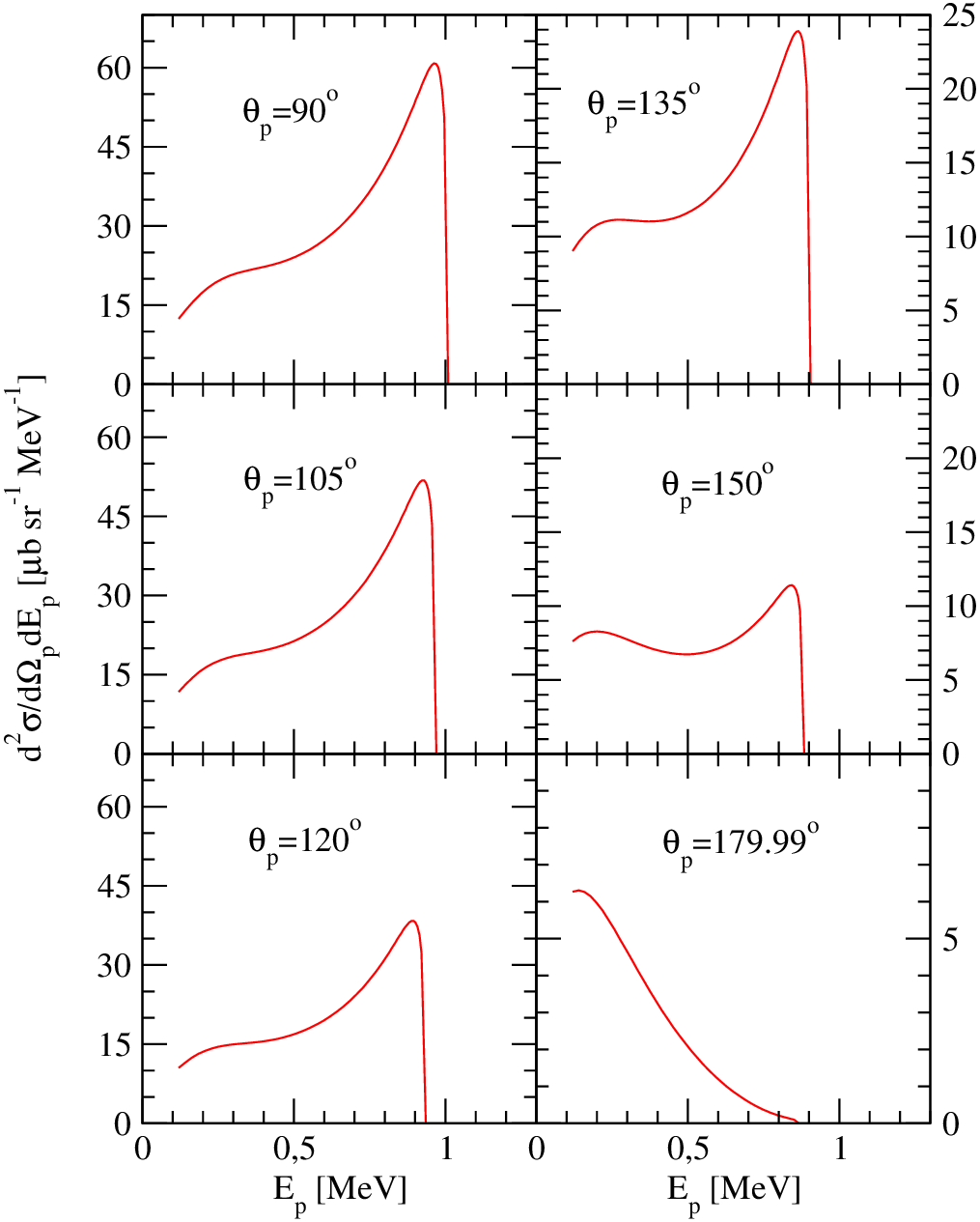}
\caption{(color online)
The spectra of the outgoing proton from the
  reaction $^3H(\gamma,p)nn$ with $E_{\gamma}=10$~MeV at different
  lab.  angles of the proton. They have been calculated using
  the 
  AV18~\cite{av18}
  NN interaction and the current composed of single nucleon and meson
  exchange currents~\cite{physrep2005}.
}
\label{fig2}
\end{figure}

It is known \cite{blank, friar}
that $^3$He is predominantely a spatially symmetric S state with its
two protons mainly in opposite spin states. This  component amounts for
$\approx 90 \%$ of the $^3$He wave function. Similarily, the two
neutrons in $^3$H are  restricted to be in a spin-singlet
state. That makes the triton target a very suitable tool to look for a nn
bound state in $\gamma$ induced breakup of $^3$H. The idea is to
measure the spectra of the outgoing protons in such a reaction. 
The two-neutron bound state, if existant, should reveal itself as a peak above
the highest available proton energy  from the 3-body decay of $^3$H. 
 We show in Figs.~\ref{fig1} to \ref{fig2}
 the outgoing proton spectra from the  $\gamma(^3H,p)nn$ reaction for a
 number of $\gamma$ energies and angles of the outgoing protons.
The big advantage of that reaction is that $\gamma$ interacts
predominantely with the proton.

Also other reactions, such as e.g. $^3$H(n,d)nn and $^3$H(d,$^3$He)nn,
 provide  conditions advantegous for two
neutrons to bind.
They are complimentary and independent from the $^3$H($\gamma$,p)nn
reaction and the data from all three processes should provide an
answer to the question whether two neutrons can form a bound state. The reaction
$^3$H(d,$^3$He)nn cannot  presently be  treated in a  theoretically
rigorous manner, however with the rapid increase in computer power such
a treatment based on Fadeev-Yakubovsky equations can be expected  in
the near future.

\section{Summary}
\label{five}

The strong discrepancy in the nn QFS nd break up configuration found 
in \cite{siepe2,ruan1} is reconsidered. It is documented 
again that at low energies
 (below $\approx 30$~MeV) the nn (np) QFS cross section depends 
dominantly only on the $^1S_0$ ($^3S_1-^3D_1$) NN force component 
and higher partial wave contributions are quite small. Furthermore 
the theoretical results are quite stable under exchange of 
the standard nuclear forces. Also the present day available 3N forces 
have negligible effect on the QFS configurations. Since no direct 
measurement of the nn force  is available there is the possibility 
that the properties of the nn force are  still unsettled. Thus simply 
multiplying the nn $ ^1S_0$
  force matrix element by a factor $ \lambda=1.08$ one can perfectly 
well reconcile theory and data. In addition we performed a more 
detailed study using the NLO chiral potential,
   which is composed of the one-pion exchange and contact 
interactions depending on  two parameters. That dependence 
allowed us to study separately variations in the scattering length
    $ a_{nn}$ leaving the effective range parameter  $ r_{eff}$
constant  
and vice versa. Thereby it turned out that the nn QFS peak height is 
very sensitive to $ r_{eff}$ and hardly sensitive to $ a_{nn}$. 
The outcome for an agreement with the  data is the requirement 
that  $ r_{eff}$ decreases  from the value $r_{eff} =2.75$~fm to 
a significantly  smaller  one, $r_{eff} = 2.41$~fm. That strongly breaks 
charge symmetry and charge independence and is not supported 
by present day chiral potential theory. 

So,  what might be a solution to remove the discrepancy?

If the data  are taken for granted there remains  the possibility 
that a di-neutron exists. We propose additional experimental 
investigations, like the $ ^3H( \gamma,p)nn$ process and evaluated 
the proton spectra at various emission angles emphasizing its high 
energy region. 

The direct inclusion of $\Delta$-degrees of freedom into $\chi$PT
allows for a rich set of additional NN and 3N force diagrams 
which are presently under investigation \cite{epel_privat1}. 
 This might reconcile theory and data also for the space-star 
discrepancy \cite{physrep96} in the nd breakup process.

Right now the situation  is unsettled.

\section*{Acknowledgments}
This work was supported by the Polish 2008-2011 science funds as the
 research project No. N N202 077435. It was also partially supported
 by the Helmholtz
Association through funds provided to the virtual institute ``Spin
and strong QCD''(VH-VI-231)  and by
  the European Community-Research Infrastructure
Integrating Activity
``Study of Strongly Interacting Matter'' (acronym HadronPhysics2,
Grant Agreement n. 227431)
under the Seventh Framework Programme of EU. 
H.W. would like to thank the Kyushu University and 
Triangle Universities Nuclear Laboratory 
 for hospitality and support during his stay in both
institutions.  
 The numerical
calculations have been performed on the
 supercomputer cluster of the JSC, J\"ulich, Germany.

\clearpage

\newpage

\end{document}